\documentclass[twocolumn,aps,prd,preprintnumbers,nofootinbib]{revtex4}
\usepackage[dvipdfmx]{graphicx}
\usepackage{bm,latexsym,amsmath,amssymb,amsfonts,color}

\usepackage[normalem]{ulem}

\begin{document} 

\preprint{KUNS-2483}

\title{\uline{}DGP braneworld with a bubble of nothing} 

\author{Keisuke Izumi${}^1$ and Tetsuya Shiromizu${}^{2,3}$}
\affiliation{${}^1$Leung Center for Cosmology and Particle Astrophysics, National Taiwan University, Taipei 10617, Taiwan}
\affiliation{${}^2$Department of Physics, Kyoto University, Kyoto 606-8502, Japan}
\affiliation{${}^3$Department of Mathematics, Nagoya University, Nagoya 464-8602, Japan}

\begin{abstract}
We construct exact solutions with the bubble of nothing in the 
Dvali-Gabadadze-Porrati braneworld model. 
The configuration with a single brane can be constructed, unlike in the Randall-Sundrum braneworld model. 
The geometry on the single brane looks like the Einstein-Rosen bridge. 
We also discuss the junction of multibranes. Surprisingly, even without any artificial matter fields on the 
branes such as three-dimensional tension of the codimension-two objects, 
two branes can be connected in certain configurations. 
We investigate solutions of multibranes too. 
The presence of solutions may indicate the semiclassical instability of the models. 
\end{abstract}
\maketitle

\section{Introduction}

The Dvali-Gabadadze-Porrati (DGP) braneworld is a model that may be able to explain the current acceleration of the Universe 
without introducing the cosmological constant~\cite{DGP}. 
Therein the four-dimensional universe is treated as a membrane with the induced gravity. The braneworld model is 
one of natural pictures of our Universe inspired by string theory, and the induced gravity is expected via 
the quantum correction into the matter fields on the brane~\cite{loop}. In the DGP models we often have the two type cosmological solutions~\cite{Deffayet,Gregory2006}, 
that is, the normal branch and the self-accelerating branch. The latter was expected to explain the  current acceleration of the 
Universe. But, it was shown that the self-accelerating branch of 
the single brane model in the DGP braneworld is not compatible with observations~\cite{Fang} and also suffers from ghost instability\footnote{
Spontaneous breaking of the local Lorentz symmetry may save the theory from the ghost disaster~\cite{Izumi:2007pb,Izumi:2008st}.
}~\cite{Koyama:2005tx,Izumi:2006ca}(see Ref. \cite{Koyama} for a review). However, 
there are still rooms for two branes models, which may realize the nonlinear massive gravity theory 
\cite{dRGT} and/or bi-gravity theory \cite{bigravity}(see Ref. \cite{deRham2014} for review) 
as an effective one \cite{Padilla}, and the normal branch for single brane model. 

In general, the spacetime with compact extra dimensions is semiclassically unstable if there is no 
fundamental fermion and/or supersymmetry. The spacetime decays to so-called Kaluza-Klein(KK) bubble-type spacetimes \cite{Witten}. 
The bubble of nothing is nucleated via the quantum gravity effect. For the four-dimensional observers, the 
spacetime is incomplete at the surface on the bubble and the surface will expand with almost light velocity. The transition rate 
from the KK vacuum to the bubble depends on the size of the initial bubble. When the size is larger than the 
Planck scale, it is exponentially suppressed. 

For the Randall-Sundrum braneworld model \cite{RS}, the similar feature was reported \cite{Ida}. 
In this paper, we discuss the same issue in the DGP braneworld context and focus on the construction of the 
braneworld model with the bubble of nothing. We will consider the normal branch 
on the brane although one may be interested in the self-accelerating branch. 
See Refs. \cite{Gregory2007, Izumi2007} for the related work (therein another decay channel was discussed, not bubble of 
nothing). 

The remaining part of this paper is organized as follows. 
In Sec.~\ref{setup}, we give the set-up for the DGP braneworld 
and the bulk spacetime. We also have a general remark. 
In Sec.~\ref{localstructure}, we derive the junction condition on the brane for the current concrete case. 
In Sec.~\ref{SecSingle}, the local embedding of branes in the bulk spacetime is discussed.
In Sec.~\ref{Sec3Djun}, we derive the condition for connecting branes.
In Sec.~\ref{SecGlobal}, we construct the spacetime globally for the single and multibranes 
cases. Finally we give the summary and discussion in Sec.~\ref{SecSum}.  


\section{Setup} \label{setup}

For simplicity, we consider the original DGP models described by the action\footnote{
Exactly say, we have to introduce the York-Gibbons-Hawking surface term \cite{York:1972sj,Gibbons:1976ue}. 
} \cite{DGP} 
%
\begin{eqnarray}
S& = & 2M^3 \int_{\rm bulk} d^5x {\sqrt {-g}}R \nonumber \\
& & +2M^3 \sum_{i}r_{i} \int_{{\rm {brane~}}i} d^4x{\sqrt {-q_i}}{}^{(4)}R(q_i) \nonumber \\
& & +\sum_{i}S_{{\rm {brane}}~i,~{\rm matter}} \ ,
\end{eqnarray}
%
where $R$ and ${}^{(4)}R$ are the five-dimensional Ricci scalar and the four-dimensional Ricci scalar of 
the branes. 
The index $i$ labels the branes.
$g_{\mu\nu}$ and $q_{i \mu\nu}$ are the metric of the bulk and the branes. $M$ is the Planck scale 
in the five dimensions. $r_i$ has a length scale. Contrasted to the conventional higher dimensional theories, the five-dimensional effect will be crucial at a larger scale than $r_i$. $S_{{\rm {brane}}~i,~{\rm matter}}$ is the action for 
the matters localized on the branes. 

Under the $Z_2$ symmetry, the junction condition is \cite{Israel}
%
\begin{eqnarray}
K_{i,\mu\nu}-K_iq_{i,\mu\nu}=r_{i}{}^{(4)}G_{\mu\nu}(q_i)-\frac{1}{2M^3}T_{i,\mu\nu}, \label{junc}
\end{eqnarray}
%
where $K_{i,\mu\nu}$ is the extrinsic curvature of the branes, 
${}^{(4)}G_{\mu\nu}(q_i)$ is the Einstein tensor for the metric $q_i$ and $T_{i,\mu\nu}$ 
is the energy-momentum tensor for the matters localized on the branes. 
The junction condition gives us the boundary condition 
for the bulk gravitational field equation, that is, the five-dimensional Einstein equation. 
The Greek indices $\{\mu,\nu,\ldots \}$ stand for the coordinate of the four-dimensional spacetime. 
Here, the unit normal vector $n^\mu$ required for the definition of the extrinsic curvature, $K_{i,\mu\nu} := q_{i,\mu}{}^\lambda\nabla_\lambda n_\nu$, is oriented to the bulk.

For simplicity, we consider the vacuum cases, $T_{i,\mu\nu}=0$. Using the Gauss equation and the Weyl tensor, 
we have the equation on the brane as 
%
\begin{eqnarray}
{}^{(4)}G_{\mu\nu}& = & r_i^2 \Bigl[\frac{2}{3}{}^{(4)}R{}^{(4)}R_{\mu\nu}-{}^{(4)}R_{\mu\alpha}{}^{(4)}R_{\nu}^{~\alpha} \nonumber \\
& &+\frac{1}{2}q_{\mu\nu}\Bigl( {}^{(4)}R_{\alpha\beta}{}^{(4)}R^{\alpha\beta}-\frac{1}{2}{}^{(4)}R^2 \Bigr)   \Bigr]-E_{\mu\nu}, \nonumber\\
&&\label{effeq}
\end{eqnarray}
%
where we have omitted $q_i$. $E_{\mu\nu}$ is the electric part of the Weyl tensor defined by 
$E_{\mu\nu}:={}^{(5)}C_{\mu \alpha \nu \beta}n^\alpha n^\beta$.
This is the DGP version of the gravitational equation on branes for the Randall-Sundrum model \cite{SMS}. 
Since the above is the quadratic equation with respect to the four-dimensional Ricci tensor, we can guess that there are 
two branches for the solutions.  When $E_{\mu\nu}=0$ and ${}^{(4)}R_{\mu\nu}(q_i)=\Lambda_i q_{i \mu\nu}$, we have 
$\Lambda_i (1-r_i^2 \Lambda_i/3)=0$, and then $\Lambda_i=0$ (normal branch) or $\Lambda_i=3/r_i^2$ (self-accelerating branch).

The bulk spacetime follows the five-dimensional vacuum Einstein equation. As the  simplest case, the bulk spacetime is 
just the five-dimensional Minkowski spacetime. In the canonical coordinate, the metric 
is $\eta_5=dy^2+\eta_4$, where $\eta_4$ is the metric of the four-dimensional Minkowski spacetime. 
The brane can be located at $y=$const. Indeed, this is a rather trivial case. 
This corresponds to the normal branch. Here we identify it with the DGP vacuum. 
The bulk metric is also written as 
the spherical Rindler coordinate $\eta_5=dz^2+z^2\gamma_4$, where $\gamma_4$ is the four-dimensional de Sitter solution 
with the positive cosmological constant of $3/r_i^2$. This belongs to the self-accelerating branch. 

In this paper we suppose that the bulk spacetime is locally identical with the KK bubble spacetimes \cite{Witten}
%
\begin{eqnarray}
ds^2=f(r)d\chi^2+f(r)^{-1}dr^2+r^2 \gamma_{ab} dx^a dx^b ,
\label{bulkmetric} 
\end{eqnarray}
%
where $f(r)=1 \mp (r_0/r)^2$ and $\gamma_{ab}$ is the metric of the three-dimensional unit de Sitter spacetime. 
This spacetime is obtained through 
the double Wick rotation of the five-dimensional Schwarzschild spacetime. 
$f(r)$ with the negative (positive) sign is corresponding to the Schwarzschild spacetime with 
the positive (negative) mass. 
The Latin indices stand for the coordinate of the three-dimensional de Sitter spacetime. 

For the KK bubble spacetime with the positive mass the periodicity 
$2\pi r_0$ for the coordinate $\chi$ makes the spacetime regular \cite{Witten}, 
while in the KK bubble spacetime with the negative mass singularity always appears at $r=0$. 
For the moment, however, we do not care about the periodicity and the singularity, because we will use the KK bubble spacetime locally.

Here we have a comment on the simplest case, that is,  
the brane is located at $\chi=$const. In this case, the extrinsic curvature of the brane vanishes. 
Therefore, 
%
\begin{eqnarray}
{}^{(4)}G_{\mu\nu}(q_i)=\frac{1}{2M^3r_{i}} T_{i, \mu\nu} \label{Einstein}
\end{eqnarray}
%
must hold on the branes. The brane metric is 
%
\begin{eqnarray}
q_i&=&f(r)^{-1}dr^2+r^2 \gamma_{ab}dx^a dx^b \nonumber\\
&=& - r^2 d \tau^2 +f(r)^{-1}dr^2+ r^2  \left(\cosh \tau \right)^2 d \Omega_2^2 ,
\end{eqnarray}
%
where $d\Omega_2^2$ is the metric of the unit sphere.
Then, the Einstein tensor on the brane is computed as 
%
\begin{eqnarray}
{}^{(4)}G^\mu{}_\nu(q_i)=\pm\frac{r_0^2}{r^4}(1,-3,1,1).
\end{eqnarray}
%
We must put the matter on the brane to be consistent with 
Eq. (\ref{Einstein}). This means that 
the energy-momentum tensor of the matter is proportional to the above. Then it is easy to 
see that the energy condition is broken. Therefore, if we suppose that the brane is at $\chi=$const, 
it is difficult to construct the physically acceptable braneworld model in the classical level. But, we may be able to realize 
it if one considers semiclassical treatment. 
This is beyond the scope of this paper.

\section{Local structure of DGP vacuum brane}\label{localstructure}

In this section, we consider the local properties of the DGP braneworld such that the 
bulk spacetimes are {\it locally} given by the KK bubble spacetime of Eq. (\ref{bulkmetric}). 
{For simplicity, we discuss vacuum branes.}
\footnote{{
In general, generic matter fields break the symmetry of $\gamma_{ab}$ in Eq. (\ref{branemet}). 
Although we can solve the trajectories of branes in principle, the analysis becomes rather complicated.}}We write down the junction condition 
on the brane to have the equation that determines the location of the brane in the bulk. 

Let us suppose that the brane is located at 
%
\begin{eqnarray}
\chi=\bar \chi_i (r).
\end{eqnarray}
%
The induced metric of the brane becomes 
%
\begin{eqnarray}
q_i=\alpha_i^{-2}dr^2+r^2 \gamma_{ab} dx^a dx^b, \label{branemet}
\end{eqnarray}
%
where
%
\begin{eqnarray}
\alpha_i:=(\bar \chi_i'^2f+f^{-1})^{-1/2} >0.\label{definitionalpha}
\end{eqnarray}
%
The normal vector to the brane is 
%
\begin{eqnarray}
n_i=\alpha_i (d\chi-\bar \chi_i' dr) , \label{defn}
\end{eqnarray}
%
and then nonzero components of the extrinsic curvature of the brane are derived as 
%
%
%
\begin{eqnarray}
K^r{}_r=\alpha_i^3\Bigl( -\frac{3}{2}\bar \chi_i' \frac{f'}{f}-\frac{1}{2}\bar \chi_i'^3 ff'-\bar \chi_i'' \Bigr)
\end{eqnarray}
%
and
%
\begin{eqnarray}
K^a{}_b=-\delta^a_b \frac{\bar \chi'_i f}{r}.
\end{eqnarray}
%

From the induced metric, the nonzero components of the Ricci tensor are 
%
%
%
\begin{eqnarray}
{}^{(4)}R^r{}_r=-3\frac{\alpha_i \alpha_i'}{r}
\end{eqnarray}
%
and
%
\begin{eqnarray}
{}^{(4)}R^a{}_b=\frac{1}{r^2}\Bigl[ 2(1-\alpha_i^2)-\alpha_i \alpha_i' r \Bigr] \delta^a_b. 
\end{eqnarray}
%
The Ricci scalar is 
%
\begin{eqnarray}
{}^{(4)}R=\frac{6}{r^2}(1-\alpha_i^2)-6\frac{\alpha_i \alpha_i'}{r}.
\end{eqnarray}
%

It is ready to consider the junction condition. The $(a,b)$ components give us 
%
%
%
\begin{eqnarray}
\bar \chi_i'=-\frac{r_i}{rf\alpha_i}(1-\alpha_i^2). \label{chieq}
\end{eqnarray}
%
The $(r,r)$ component implies 
%
\begin{eqnarray}
& & \alpha_i^3 \Bigl(\frac{3}{2} \bar \chi_i' \frac{f'}{f}+\frac{1}{2}\bar \chi_i'^2 ff'+\bar \chi_i''  \Bigr) \nonumber \\
& & ~~=r_{i} \Bigl(2\frac{\alpha_i \alpha_i'}{r}+\frac{1}{r^2}(1-\alpha_i^2) \Bigr).
\end{eqnarray}
%
Note that this must be automatically satisfied when Eq. (\ref{chieq}) holds, 
because they are related through the energy conservation law on the brane (for example, see Ref. \cite{Sasaki:1999mi}). 

Together with the definition of $\alpha_i$, Eq. (\ref{chieq}) gives us two solutions of $\alpha_i^2$ as 
%
\begin{eqnarray}
\alpha_{\pm,i}^2=\frac{-1+2(r_{i}/r)^2 \pm{\sqrt {1-4(r_0/r)^2(r_{i}/r)^2}}}{2(r_{i}/r)^2} \label{alphap}
\end{eqnarray}
%
for the positive mass case and
%
\begin{eqnarray}
\alpha_{\pm,i}^2=\frac{-1+2(r_{i}/r)^2 \pm{\sqrt {1+4(r_0/r)^2(r_{i}/r)^2}}}{2(r_{i}/r)^2} \label{alphan}
\end{eqnarray}
%
for the negative mass case.
In the limit $r\to\infty$, $\alpha_{+,i}$ approaches unity, while $\alpha_{-,i}$ does not have solutions. 
It means that the brane with $\alpha_{+,i}$ has asymptotically flat structure and thus the same 
asymptotic structure as that of the normal branch for the Minkowski bulk. 
On the other hand, the brane with $\alpha_{-,i}$ does not exist in the asymptotic region 
(although it could exist within certain finite $r$). 


\section{Trajectories of Single Brane}
\label{SecSingle}

The number of solutions of $\alpha_i$ depends on the ratio of $r_i$ to $r_0$, which stems from requiring the 
presence of the square root in Eq. (\ref{alphap}) or (\ref{alphan}) and the positivity of $\alpha_i^2$. 
Below we will discuss the four cases (A)-(D), separately. 
Most of the trajectories of the brane in the ($r,\chi$) plane of the bulk terminate with the finite length or hit singularity. 
Only a trajectory for $r_0>r_i$ with the positive mass bulk goes to infinity, 
and thus, it would be geodesically complete. 
In this section, however, we do not care about the incompleteness of the trajectories and the singularities 
at the ``edge" of the trajectories. The purpose in this section is deriving all possibilities of the local 
embedding of branes in the bulk spacetime with the metric of Eq. (\ref{bulkmetric}). 

Hereafter we call the brane satisfying Eq. (\ref{alphap}) or (\ref{alphan}) for $\alpha_{+,i}$ ($\alpha_{-,i}$) 
the $+(-)$brane regardless of the positive or negative mass bulk. 

\subsection{$r_0>r_i$ in positive mass bulk} \label{single}

The presence of the square root in Eq. (\ref{alphap}) implies the condition $r\ge\sqrt{2r_0r_i}$ on the brane. 
$\alpha_{+,i}^2$ is positive only for $r \ge r_{*,i}:=\sqrt{r_0^2+r_i^2}$, while 
$\alpha_{-,i}^2$ always  becomes negative. 
As a result, only $+$branes can exist in the range $r \ge r_{*,i}$.

\begin{figure}
\includegraphics[width=8cm,clip]{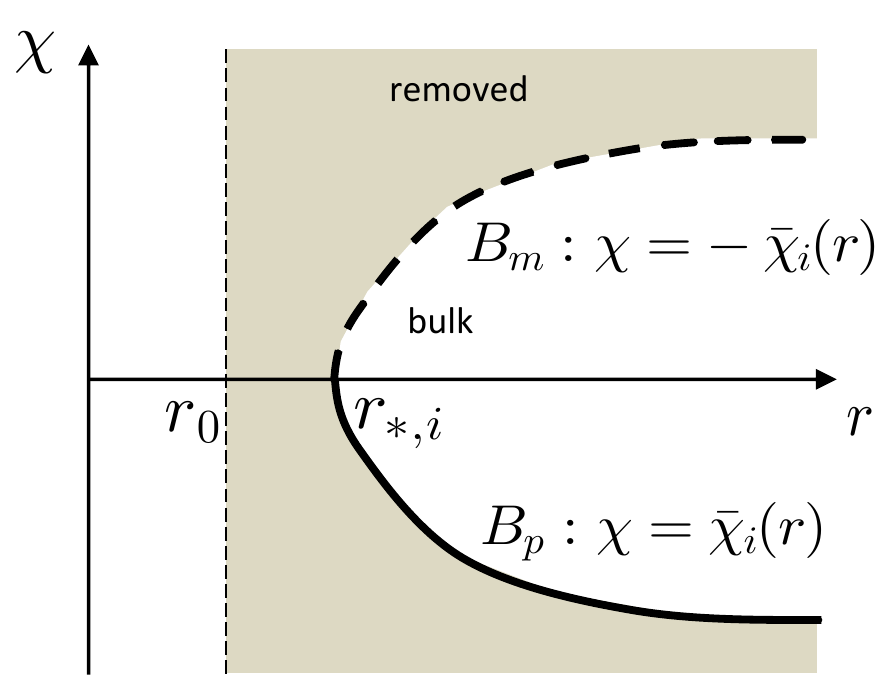}
\caption{The location of the $+$brane in the ($\chi,r$) plane for $r_0>r_i$ and the positive mass bulk:  
The gray region is removed. This is only a regular solution with the single brane.}
\label{fig1}
\end{figure}

The bulk (and the forbidden region) can be fixed by the direction of 
the unit normal vector $n_\mu$ as commented below Eq. (\ref{junc}). 
The unit normal vector $n_\mu$ is defined in Eq. (\ref{defn}) and 
the coefficient of $dr$ is $-\alpha_i {\bar \chi_i}'$. 
From the definition of $\alpha_i$, i.e. Eq. (\ref{definitionalpha}), $\alpha_i$ must be smaller than unity. 
Combining this result with Eq. (\ref{chieq}), it is easy to show the positivity of $-\alpha_i {\bar \chi_i}'$. 
This means that the  unit normal vector $n_\mu$ is pointing in the direction of increasing $r$, and thus, 
the region, where the coordinate $r$ is smaller than that on the brane with the same value of $\chi$, is 
forbidden and the remaining region becomes bulk (see Fig.~\ref{fig1}). 
At the brane, the $Z_2$ symmetry is imposed.

By the integration of Eq. (\ref{chieq}) with the boundary condition ${\bar \chi_i}(r_{*,i})=0$, we can obtain the trajectory of a $+$brane.
However, the surface of $\chi={\bar \chi_i}(r)$, say $B_p$, 
is incomplete at $r=r_{*,i}$. This can be geodesically complete by reflecting with respect to 
the $\chi=0$ surface. The sum with the reflected surface $B_m$ of $\chi=-{\bar \chi_i} (r)$, $B=B_p \cup B_m$, is geodesically complete. The bulk spacetime is the region of $r \geq r_0$ removing the gray region as Fig.~\ref{fig1}.  

\subsection{$r_0<r_i$ in positive mass bulk}

\begin{figure}
\includegraphics[width=8cm,clip]{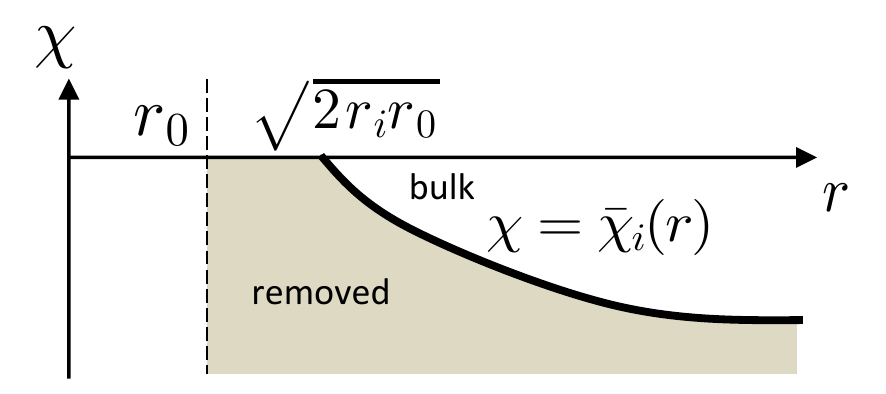}
\caption{The location of the $+$brane in the ($\chi,r$) plane for $r_0<r_i$ and the positive mass bulk: 
The gray region is removed. The trajectory of the brane cannot be extended beyond $r=\sqrt{2r_ir_0}$}
\label{fig2}
\end{figure}

The presence of the square root in Eq. (\ref{alphap}) constrains a lower limit of $r$  
on the brane as $r\ge\sqrt{2r_0r_i}$. The positivity of $\alpha_i^2$ implies the upper limit $r_{*,i}={\sqrt {r_0^2+r_i^2}}$ 
only for $\alpha_{-,i}$. As a result, the $+$brane is embedded in the range $\sqrt{2r_0r_i}\le r$ and the $-$brane is in the range 
$\sqrt{2r_0r_i}\le r \le r_{*,i}$. 
The argument of choosing the bulk region is the same as that in the previous case (see Fig. \ref{fig2}). 

\begin{figure}
\includegraphics[width=8cm,clip]{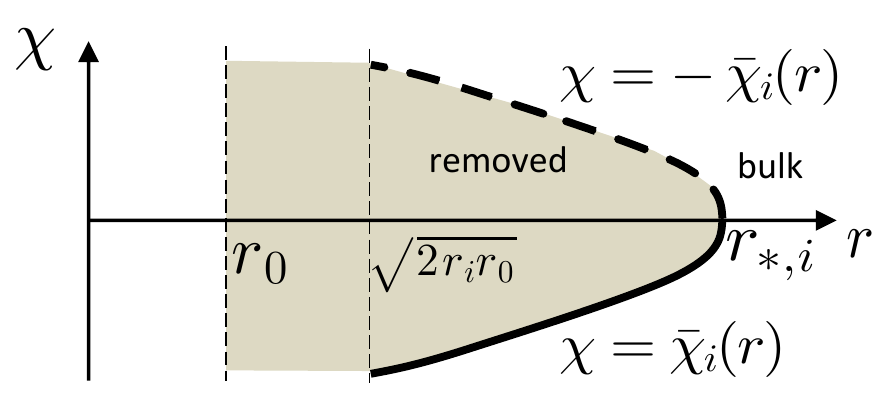}
\caption{The location of the $-$brane in the ($\chi,r$) plane for $r_0<r_i$ and the positive mass bulk: 
The gray region is removed. The trajectory of the brane cannot be extended to the region of $r \leq \sqrt{2r_ir_0}$.}
\label{fig3}
\end{figure}

Unlike in the previous case, the $+$brane cannot be smoothly connected with its reflected image at the minimum value of $r$, ${\sqrt {2r_0r_i}}$. 
On the other hand, we can connect a $-$brane with its reflected image at $r=r_{*,i}$ in the same way of the previous case for $+$branes. 
Then, we have two possible solutions, on both branes of which geodesics are incomplete at $r=\sqrt{2r_0r_i}$
(see Figs.~\ref{fig2} and \ref{fig3}). 

\subsection{$r_0<r_i$ in negative mass bulk}

The square root in Eq. (\ref{alphan}) is always positive, and thus, there are no restrictions for the range of $r$. 
The positivity of $\alpha_i^2$ gives the upper limit ${\bar r}_{*,i}:= \sqrt{r_i^2-r_0^2}$ only for the $-$brane. 
The bulk geometry has a singularity at $r=0$, and all single branes touch the singularity. 
As in the previous case, only the $-$brane can be connected with its reflected image at $r={\bar r}_{*,i}$. 

\begin{figure}
\includegraphics[width=8cm,clip]{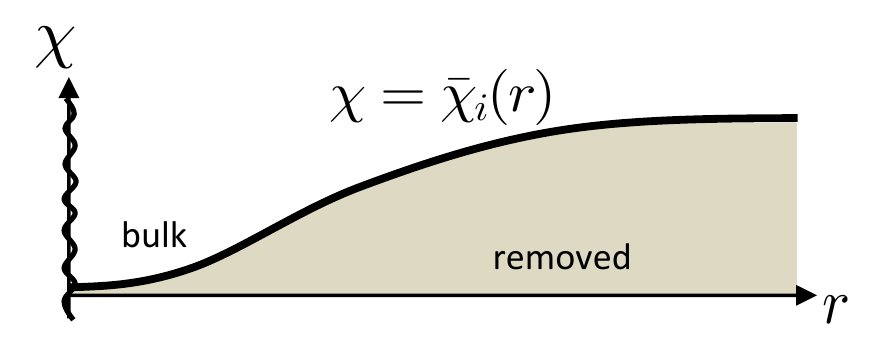}
\caption{The location of the $+$brane in the ($\chi,r$) 
plane for the negative mass bulk: The gray region is removed. The trajectory of the brane hits the singularity at $r=0$.}
\label{fig4}
\end{figure}

\begin{figure}
\includegraphics[width=8cm,clip]{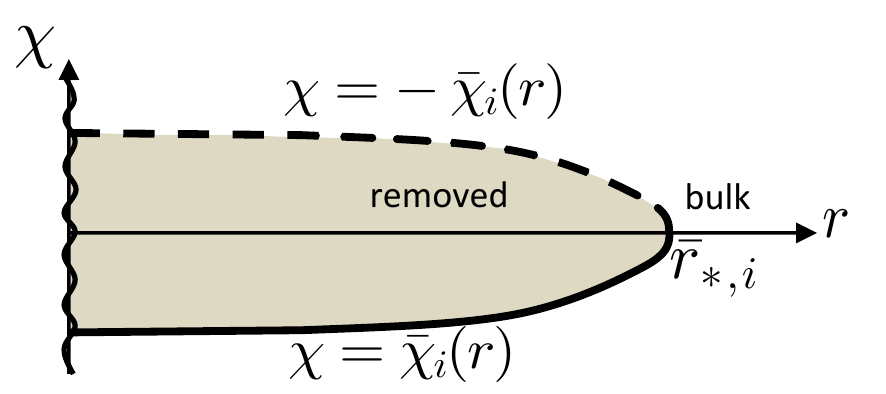}
\caption{The location of the $-$brane in the ($\chi,r$) plane for the negative mass bulk: 
The gray region is removed. The trajectory of the brane hits the singularity at $r=0$.}
\label{fig5}
\end{figure}

The bulk region for the $\pm$branes is determined as shown in Figs. \ref{fig4} and \ref{fig5}. 
Here we note that $\alpha_{+,i}^2$ always becomes larger than unity, which can be directly seen from Eq. (\ref{alphan}), and 
this makes the sign of $(1-\alpha_i^2)$ flipped. Then the position of bulk (Fig. \ref{fig4}) appears on the opposite side compared to the 
$-$brane case (Fig. \ref{fig5}). 

\subsection{$r_0>r_i$ in negative mass bulk}

For $+$branes, the discussion is the same as the previous one. 
Meanwhile, $\alpha_{-,i}^2$ always becomes negative, and thus, the $-$branes configuration does not exist. 
As a result, we have only a +branes configuration (see Fig.~\ref{fig4}).

%
%

\section{Brane junction}
\label{Sec3Djun}

In this section we shall discuss the possible brane junctions locally. 
In general, several branes intersect each other. At the junctions 
between branes, there is a restriction from field equations. 
In this section, we derive the equations for that. 

We perform the integration of the equation for the vicinity of the 
junction point and then take the limit such that the integration 
domain goes to zero, as in the derivation of the junction condition 
for singular surfaces \cite{Israel}. The integration of the $(a,b)$ component of the five-dimensional Einstein equation with respect to $\chi$ and $r$ gives 
\begin{eqnarray}
& & 2M^3 \int_{{\rm {bulk}}} d\chi dr \  G^a{}_b 
 +2M^3 \sum_{i}r_{i} \int_{{\rm {brane~}}i} dr \  {}^{(4)}G(q_i)^a{}_b \nonumber \\
& & \qquad\qquad\qquad
-\frac{1}{2} \sum_{i} \int_{{\rm {brane~}}i} dr \ T_{i,}{}^a{}_b=0,
\label{3Djunc}
\end{eqnarray}  
where $G^a{}_b$ is the five-dimensional Einstein tensor.

\begin{figure}
\includegraphics[width=8cm,clip]{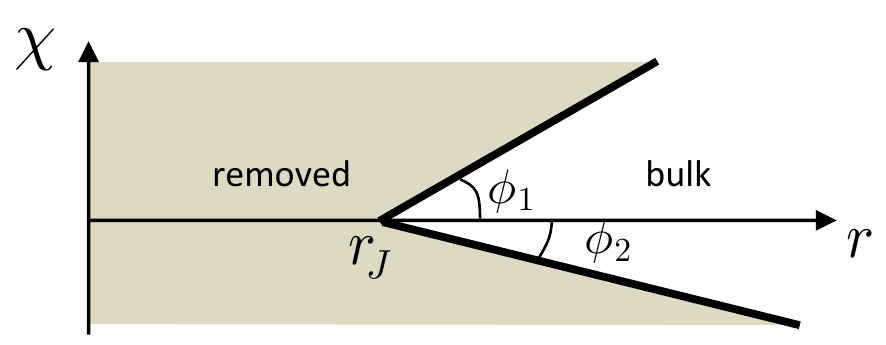}
\caption{Case 1: Both branes approach the junction point $r=r_J$ 
from larger $r$.}
\label{case1}
\end{figure}

\begin{figure}
\includegraphics[width=8cm,clip]{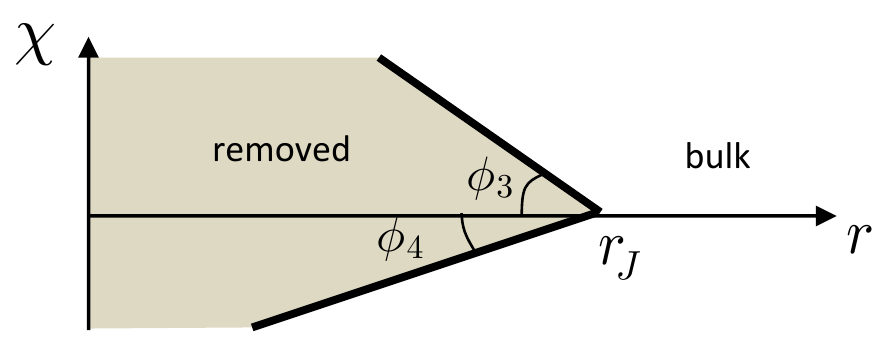}
\caption{Case 2: Both branes approach the junction point $r=r_J$ 
from smaller $r$.}
\label{case2}
\end{figure}

\begin{figure}
\includegraphics[width=8cm,clip]{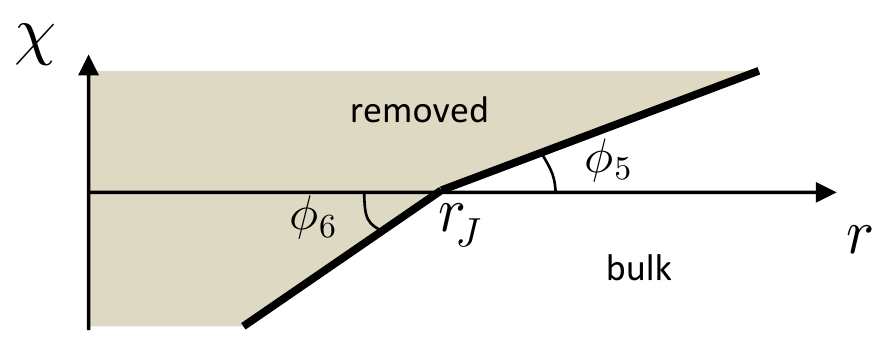}
\caption{Case 3: One brane approaches the junction point $r=r_J$ 
from larger $r$, while another approaches the point $r=r_J$ from smaller 
$r$. }
\label{case3}
\end{figure}

\begin{figure}
\includegraphics[width=8cm,clip]{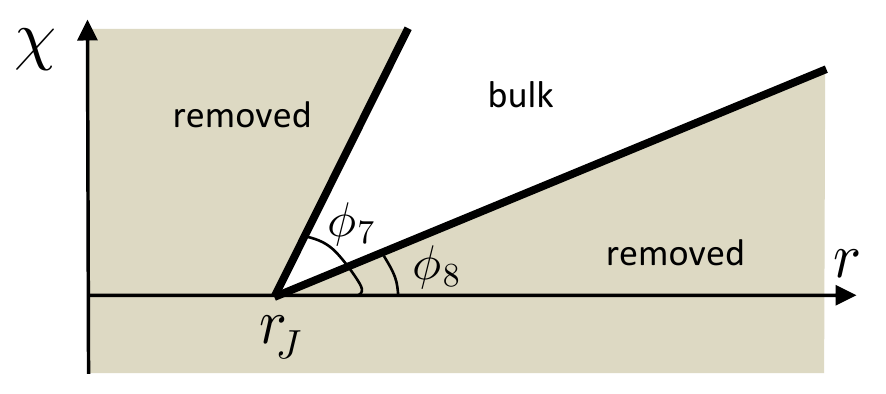}
\caption{Case 4: The $-$brane is connected with a $+$brane for negative mass bulk. }
\label{case4}
\end{figure}

We classify the brane junctions into the four cases (Figs.~\ref{case1}-\ref{case4}). 
The first three, Figs.~\ref{case1}-\ref{case3}, are  brane junctions in the positive mass bulk or 
$-$brane junctions in the negative mass bulk, where $\alpha_i$ is always smaller than unity. 
The last one, Fig.~\ref{case4}, describes the junctions of the $+$brane and the $-$brane in the negative mass bulk. 
We will look at them in  detail. 

\subsection{Contribution from five-dimensional bulk gravity}

The first term of Eq. (\ref{3Djunc}) is evaluated through the contribution 
from the deficit angle $\phi$ as~\cite{Israel:1976vc}
\begin{eqnarray}
\int dx^2 \  G^a{}_b = - \frac{\phi}{2} \delta^a_b. \label{deficit1}
\end{eqnarray}
Therefore, what we have to do is only deriving the deficit angle. Then we 
compute it for each case. 
%

\begin{figure}
\includegraphics[width=8cm,clip]{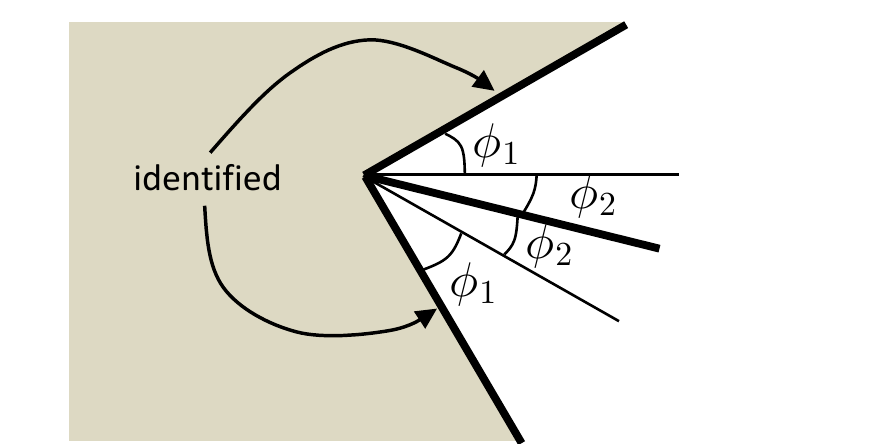}
\caption{Gluing the mirror image due to $Z_2$-symmetry: 
The deficit angle becomes $2\pi-2\phi_1-2\phi_2$.}
\label{copy}
\end{figure}

(i) In case 1 (Fig.~\ref{case1}), both branes go to the direction of 
increasing $r$ from the junction point. Since the $Z_2$ symmetry is imposed across 
the branes, we can construct the bulk locally as Fig.~\ref{copy}. 
Then, the deficit angle is estimated at $2\pi - 2\phi_1- 2\phi_2$, where 
$\phi_1$ and $\phi_2$ are defined in Fig.~\ref{case1} and they are taken to be 
a smaller value than $\pi$. 

Using the bulk metric (\ref{bulkmetric}), the angle $\phi_i$ can be written as 
\begin{eqnarray}
\phi_i&=&\arctan \left| \frac{f^{1/2} }{f^{-1/2} } \frac{d\chi}{dr} \right| _{r=r_J} \nonumber\\
&=& \left. \arctan \left(\frac{r_i}{r}\frac{1-\alpha_i^2}{\alpha_i}\right)\right|_{r=r_J},
\label{deficit2}
\end{eqnarray}
where the branes are connected at $r=r_J$. Finally, from Eqs. 
(\ref{deficit1}) and (\ref{deficit2}), we see
\begin{eqnarray}
\int d\chi dr \  G^a{}_b 
=\left. \sum_{i=1,2}\left[ -\frac{\pi}{2} + \arctan \left(\frac{r_i}{r}\frac{1-\alpha_i^2}{\alpha_i}\right) \right] \right|_{r=r_J}\!\!\!\!\!\!\!\!\!\!\!\! \delta^a_b. \nonumber\\
\label{deficit}
\end{eqnarray}

%

(ii) For case 2 (Fig.~\ref{case2}), both branes go to the direction of decreasing $r$ from the junction point.
We can see the deficit angle becomes 
\begin{eqnarray}
2\pi-2(2\pi-\phi_3-\phi_4)=-2\pi+2\phi_3+2\phi_4.
\end{eqnarray}
Then, 
\begin{eqnarray}
\int d\chi dr \  G^a{}_b 
=\left. \sum_{i=3,4} - \left[ -\frac{\pi}{2} + \arctan \left(\frac{r_i}{r}\frac{1-\alpha_i^2}{\alpha_i}\right) \right] \right|_{r=r_J}\!\!\!\!\!\!\!\!\!\!\!\! \delta^a_b. \nonumber\\
\end{eqnarray}
%

(iii) In  case 3 (Fig.~\ref{case3}), the branes go to the opposite directions from each other with respect to $r$. 
In the same method the deficit angle becomes 
\begin{eqnarray}
2\pi-2(\pi+\phi_5-\phi_6)=-2\phi_5+2\phi_6,
\end{eqnarray}
and then
\begin{eqnarray}
\int d\chi dr \  G^a{}_b 
=\left. \sum_{i=5,6} \epsilon_i \left[ -\frac{\pi}{2} + \arctan \left(\frac{r_i}{r}\frac{1-\alpha_i^2}{\alpha_i}\right) \right] \right|_{r=r_J}\!\!\!\!\!\!\!\!\!\!\!\! \delta^a_b, \nonumber\\
\end{eqnarray}
where $\epsilon_i$ is unity for $\phi_5$ and $-1$ for $\phi_6$. 
%
 
(iv) In  case 4 (Fig.~\ref{case4}), both branes go to the direction of 
increasing both $\chi$ and $r$ from the junction point. Since $|{\bar \chi_i}'|$ for 
the $-$brane is larger than that for the $+$brane, the $-$brane corresponds to 
one with the angle $\phi_7$ in Fig.~\ref{case4}. 
The deficit angle is $2\pi-2(\phi_7-\phi_8)$. Here note 
that the sign of $(1-\alpha^2)$ for the $+$brane is different from that 
in the previous cases, and then we compute as 
\begin{eqnarray}
\phi_8&=&\arctan \left| \frac{f^{1/2} }{f^{-1/2} } \frac{d\chi}{dr}\right|_{r=r_J} \nonumber\\
&=& \left. - \arctan \left(\frac{r_8}{r}\frac{1-\alpha_8^2}{\alpha_8}\right)\right|_{r=r_J}.
\end{eqnarray}
For $\phi_7$, we can use Eq. (\ref{deficit2}), and then we arrive at 
Eq. (\ref{deficit}). 

\subsection{Contribution from four-dimensional induced gravity}

The second term of Eq. (\ref{3Djunc}) comes from the discontinuity of the first derivative of 
the induced metric on the brane. Without loss of generality, we use the Gaussian normal 
coordinate $\bar r$ on the brane
\begin{eqnarray}
d \bar r= \frac{dr}{\alpha_i}.
\end{eqnarray}
Then, the induced metric on the brane is written as 
\begin{eqnarray}
q_i= d {\bar r}^2 + r^2(\bar r) \gamma_{ab} dx^a dx^b.
\end{eqnarray}
On the brane, the extrinsic curvature $H_{ab}$ of the $\bar r=$const 
surfaces is given by 
\begin{eqnarray}
H_{ab}&:=& \frac{1}{2} \frac{\partial}{\partial \bar r} \left( r^2(\bar r) \gamma_{ab} \right)
\nonumber\\
&=& \frac{\alpha_i}{r(\bar r)}r^2(\bar r) \gamma_{ab}.
\end{eqnarray}
Since ${}^{(4)}G^a{}_b=-\partial_{\bar r}H^a{}_b+\partial_{\bar r}H^c{}_c \delta^a_b+\cdots$, 
the integration of the four-dimensional gravity term becomes 
\begin{eqnarray}
\sum_{i}r_{i} \int_{{\rm {brane~}}i} dr \  {}^{(4)}G(q_i)^a{}_b
= \left. \sum_{i} 2 \epsilon_i \frac{r_{i}}{r}\alpha_i \right|_{r=r_J}
\!\!\!\!\!\!\!\!\!\!\!\! \delta^a_b ,
\label{cont4D}
\end{eqnarray}
where $\epsilon_i=1$ if the brane goes from the junction point to 
the direction of increasing $r$ (e.g. both branes in Fig.~\ref{case1}) 
and $\epsilon_i=-1$ with the opposite direction. 

\subsection{Condition for brane junction}

Now we are ready to derive the explicit form of the condition Eq. (\ref{3Djunc}). 
Since the first and second terms in Eq. (\ref{3Djunc}) 
are proportional to $\delta^a_b$, 
the energy-momentum tensor of matter $T_{i,}{}^a{}_{b}$, if it exists, should be so.
Thus, we introduce only three-dimensional tension: 
\begin{eqnarray}
T_{i,}{}^a{}_b=-\mu_i \delta^a_b.
\end{eqnarray}
Summing up all, finally we obtain the junction condition,
\begin{eqnarray}
\sum_i \left[ \epsilon_i h_i(r_J) + \frac{\mu_i}{4M^3}\right]=0,
\label{3Djun}
\end{eqnarray}
with
\begin{eqnarray}
h_i(r) := -\frac{\pi}{2} + \arctan \left(\frac{r_i}{r}\frac{1-\alpha_i^2}{\alpha_i}\right) 
+2  \frac{r_{i}}{r}\alpha_i.
\end{eqnarray}

\section{Global Solutions} \label{SecGlobal}

In this section, we construct the global solutions 
that are asymptotically flat on the branes. 
The simplest one is the single brane configuration discussed in 
Sec.~\ref{single}~A. 
Moreover, if we consider the junction of two or multibranes, we can 
construct many nontrivial configurations. 
For instance, by considering the junction of two $+$branes, we can construct 
the configurations where the induced metric on the branes approaches flatness at 
both asymptotic regions (see Fig.~\ref{twobrane1}).  
Another asymptotically flat brane configuration is achieved by 
connecting two $+$branes with $-$branes as shown in Fig.~\ref{twobrane2}. 
At each junction, Eq. (\ref{3Djun}) must be satisfied. 
Generically we need the three-dimensional tension, i.e. the energy momentum 
tensor of the domain wall on the branes. For certain configurations, however, the three-dimensional tension is absent. 
This happens when the contribution from the five-dimensional gravity to the 
deficit angle is balanced with that from the four-dimensional gravity. This is significant 
difference from the Randall-Sundrum model where the balance does not work. 

\subsection{Single brane case}

The simplest global solution accommodated with the asymptotically flat 
condition is that with a single brane discussed in Sec.~\ref{single}~A. 
The condition $r_0>r_i$ is required for the guarantee of the existence 
of the global solution. 
This means that the bulk spacetime contains a large bubble of nothing. 
Note that the minimum size of the bubble is $r_{*,i}$, which is larger than 
$r_0$. 

We shall discuss the geometry on the brane shortly. 
Introducing the null coordinates $u_\pm$ defined by $d u_\pm =d\tau \pm dr/(r \alpha_i)$, 
the induced metric is written as 
%
\begin{eqnarray}
q_i=-r^2du_+du_-+{\cal R}^2(u_+,u_-)d\Omega_2^2,
\end{eqnarray}
%
where ${\cal R}=r \cosh \tau$. The expansion of null is given by 
%
\begin{eqnarray}
\theta_\pm=\frac{\partial \ln {\cal R}}{\partial u_\pm}=\frac{1}{2}(\pm \alpha_i+\tanh \tau).
\end{eqnarray}
%
We see that $\theta_+$ or $\theta_-$ vanishes at 
%
\begin{eqnarray}
r^2(\tau)=\frac{r_c^2}{\cosh^2 \tau}+r_0^2 \cosh^2 \tau. 
\label{expansion}
\end{eqnarray}
%
Since the right-hand side of the above equation is larger than or equal to 
$r_{*,i}^2$ for $r_0>r_c$, the solution to the above always exists. 
Moreover, it is easy to show that the hypersurface ${\cal H}$ specified by the above is timelike. 
Along the hypersurface, $\theta_+=0, \theta_-=\tanh \tau \leq 0$ for $\tau \leq 0$ and 
$\theta_+ =\tanh \tau \geq 0, \theta_-=0$ for $\tau \geq 0$. Note that $\theta_\pm=0$ at $\tau=0$. 
Therefore, ${\cal H}$ is like the apparent horizon for $\tau<0$ and the cosmological horizon for $\tau>0$.  
{The brane has two asymptotically flat regions, and then we see that}
the geometry is like the Einstein-Rosen bridge and is similar with that in 
the Randall-Sundrum models  with a bubble of nothing \cite{Ida}.%
\footnote{Solutions in which the bulk geometry is like a wormhole have been discussed 
in Ref.~\cite{Richarte:2010bd,Richarte:2013lua}.} 
 {Since we consider the vacuum brane in the DGP braneworld model, all of the dominant, null and weak energy 
conditions are trivially satisfied. Meanwhile, one may want to regard the right-hand side of Eq. (\ref{effeq}) as 
the effective energy-momentum tensor. It is easy to see that it does not satisfy all of the energy conditions.}


\subsection{Multibranes case}

\begin{figure}
\includegraphics[width=8cm,clip]{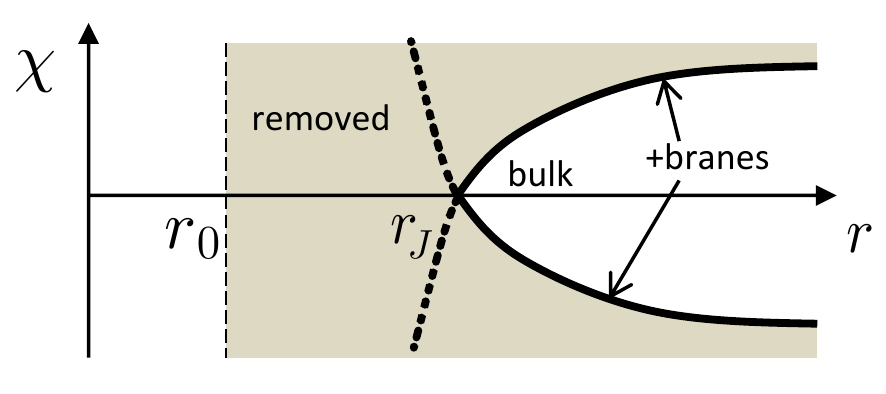}
\caption{Global configuration composed of two $+$branes.}
\label{twobrane1}
\end{figure}

\begin{figure}
\includegraphics[width=8cm,clip]{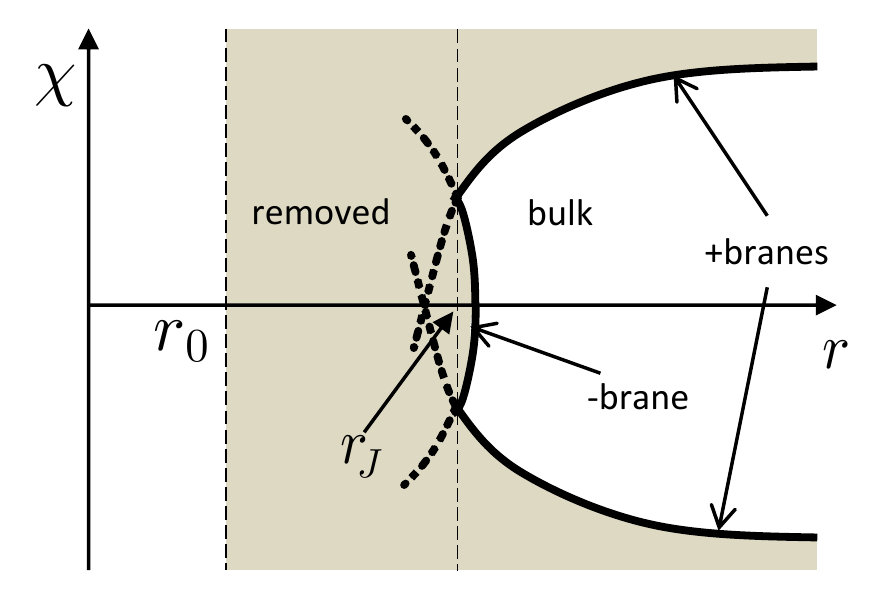}
\caption{Global configuration composed of two $+$branes with a single $-$brane: Two +branes are connected with a $-$brane at $r=r_J$. In the local aspect of the junctions, this case 
belongs to case 1 in Sec.~\ref{Sec3Djun}.}
\label{twobrane2}
\end{figure}

\begin{figure}
\includegraphics[width=8cm,clip]{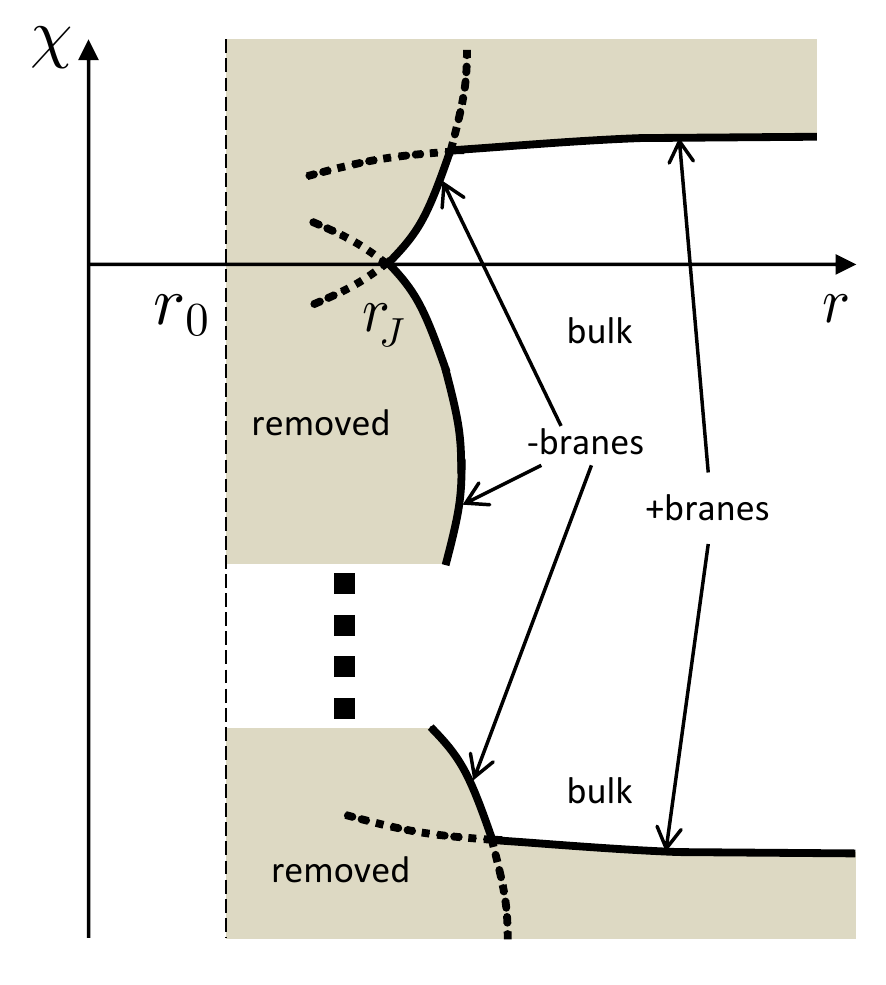}
\caption{Global configuration composed of two $+$branes with multi 
$-$branes. The junction here belongs to  cases 1 and 3 
in Sec.~\ref{Sec3Djun}.}
\label{twobrane3}
\end{figure}

We investigate the possibility to connect branes with and without tension 
terms of codimension-two object. 
For simplicity, we consider the cases where all branes have the same $r_i$, say $r_c$. 
Here, we concentrate on the three cases: (a) two $+$branes (Fig.~\ref{twobrane1}), 
(b) two $+$branes with a single $-$brane (Fig.~\ref{twobrane2}) 
and (c) two $+$branes with multi $-$branes (Fig.~\ref{twobrane3}).  
We call $h_i$ with $\alpha_{+,i}$ ($\alpha_{-,i}$) $h_+$ ($h_-$).

For later convenience, we note that $h_i(r)$ is a monotonically decreasing 
function with respect to $r$. 
Using Eqs. (\ref{alphap}) and (\ref{alphan}), indeed, 
we can derive 
\begin{eqnarray}
&&\frac{d h_i(r)}{dr} =\nonumber\\
&&- \frac{r_i \left[\alpha_i^2\left\{r^2 \left(1+\alpha_i^2\right)+2r_i^2\left(1-\alpha_i^2\right)^2\right\} +r^2 \left(1-\alpha_i^2\right)^2\right]}{r^2 \alpha_i \left( r^2 \alpha_i^2 +r_i^2 (1-\alpha_i^2)^2\right)} \nonumber \\
&&\qquad\qquad\qquad\qquad\qquad\qquad\qquad\qquad\qquad\qquad
<0.
\end{eqnarray}

\subsubsection*{\textsl{(a)} Two $+$branes }

This configuration is possible only for a positive mass bulk.  From the definition it is easy to see that 
$h_+ (r)$ approaches $-\pi/2$ in the limit $r\to \infty$. 

For $r_0>r_c$, the possible minimum value of $r_J$ is $r_*:=\sqrt{r_c^2+r_0^2}$, and $h_+ (r_*)$ becomes zero. 
At the point $r_J$ satisfying $h_+ (r_J)=0$, we see from Eq. (\ref{3Djun}) that two branes can be connected without 
introducing tension $\mu_i$. 
However, the connection at $r=r_*$ becomes regular, and it is nothing but a single brane given in SEc.~\ref{single}~A. 
For $r_J>r_*$, $h_+(r_J)$ becomes negative because of its monotonically decreasing feature, 
and thus, we need to introduce codimension-two object with positive tensions to be consistent with Eq. (\ref{3Djun}). 

Next, we consider the cases of $r_0<r_c$. We will ask if there is a case such that we can construct nontrivial configurations 
without introducing the codimension-two object with tension. To do so we will examine the existence of $r_t$ such that 
$h(r_t)=0$. We first evaluate $h_+(r)$ at $r=r_*$ 
\begin{eqnarray}
h_+(r_*)= -\frac{\pi}{2}+ \arctan\left(\frac{r_0^2}{\sqrt{r_c^4-r_0^4}}\right) 
+ 2\sqrt{\frac{r_c^2-r_0^2}{r_c^2+r_0^2}}.
\end{eqnarray}
We can show the positivity of this. Introducing the parameter $y$ defined by
\begin{eqnarray}
r_0^2=r_c^2-y^2,~(0<y<r_c) 
\end{eqnarray}
and regarding $h_+(r_*)$ as the function of $y$, $F(y)$, we see 
\begin{eqnarray}
\frac{d F(y)}{dy} = \frac{2 \left(\cos\theta_y\right)^2 r_c^4}{\left(2r_c^2-y^2\right)^\frac{5}{2}} >0, \label{dFdy}
\end{eqnarray}
where
\begin{eqnarray}
\theta_y := \arctan\left(\frac{r_0^2-y^2}{\sqrt{r_c^4-r_0^4}}\right) .
\end{eqnarray}
Since $F(0)=0$, the above tells us the positivity of $F(y)$, that is, $h_+(r_*)>0$. 
Because in the asymptotic region (i.e. large $r$) $h_+ (r)$ become negative, 
there is the point $r=r_{t}>r_*$ such that $h_+ (r_{t})=0$. 
At $r=r_{t}$, therefore, we can connect two $+$branes without tension terms. 
If the junction point $r_J$ is $r_J>r_{t}$, we need positive tension terms to connect two $+$branes, while 
negative tension terms are needed if $\sqrt{2r_0r_c}<r_J<r_{t}$.

\subsubsection*{\textsl{(b)} Two $+$branes with a single $-$brane}

This configuration 
is possible only in the case with $r_0<r_c$. 

Since the $-$brane can be in the range $\sqrt{2r_cr_0}<r<r_*$ for the positive mass, 
branes should be connected in this region.
We can easily obtain $h_-(r_*)=0$ while we saw $h_+(r_*)>0$. 
Since $h_i (r)$ is a monotonically decreasing function of $r$, $h_+(r)+h_-(r)$ 
is always positive for $\sqrt{2r_cr_0}<r<r_*$. 
Therefore, we see from Eq. (\ref{3Djun}) that only negative tension terms can make the branes connected. 

For the negative mass bulk, the situation is similar to the positive mass case. First of all, it is easy to see 
$h_-(\bar r_*)=0$ through the direct calculation, where $\bar r_*=\sqrt{r_c^2-r_0^2}$. 
The value of $h_+(\bar r_*)$ is written as
\begin{eqnarray}
h_+(\bar r_*)= -\frac{\pi}{2}+ \arctan\left(\frac{ - r_0^2}{\sqrt{r_c^4-r_0^4}}\right) 
+ 2\sqrt{\frac{r_c^2+r_0^2}{r_c^2-r_0^2}}.
\end{eqnarray}
Introducing the parameter $y$ as 
\begin{eqnarray}
r_0^2=y^2- r_c^2, \quad \mbox{with} \quad r_c<y<\sqrt{2} r_c,
\end{eqnarray}
we regard $h_+(\bar r_*)$ as the function of $y$ as 
\begin{eqnarray}
h_+(\bar r_*)= F(y).
\end{eqnarray}
Here note that $F(y)$ is the same as that introduced before. 
Since we have already shown the positivity of $F(y)$ for $y>0$, $h_+(\bar r_*)$ is positive. 
Then the monotonically decreasing property of $h_i (r)$ implies the positivity of $h_+(r)+h_-(r)$ for $0<r<\bar r_*$, and 
Eq. (\ref{3Djun}) shows us that the negative tension terms are needed to connect the branes. 

As a result, for this configuration of branes, we need a negative tension term in this configuration. 
It is probably unphysical because of the negative energy density. 

\subsubsection*{\textsl{(c)} Two $+$branes with multi $-$branes}

This configuration
is possible only for $r_0 <r_c$. 
We can consider both cases of positive and negative mass bulks. 
This configuration always has the junction between two $-$branes. 
Let us look at the details shortly. 

$h_-(r)$ becomes zero at $r=r_*$ for a positive mass bulk and at $r=\bar r_*$ for a negative mass bulk. 
The monotonically decreasing property of $h_-(r)$ leads to the positivity of $h_-(r)$.  
As a result, it is impossible to construct physically interesting solutions without introducing 
negative tension terms.

\section{summary and discussion}
\label{SecSum}

In this paper we constructed the DGP braneworld with a bubble of nothing. Surprisingly, we could 
have the single brane solutions. 
This is impressive because we could not for the Randall-Sundrum 
braneworld. 
The solution with a single brane exists only for $r_0>r_i$, while for $r_0<r_i$ solutions 
with connected two branes can be constructed even without any matter fields on branes. 
Therein, the contribution of deficit structure on five-dimensional spacetime 
is balanced with that of a singular surface on the brane, that is, codimension-two objects in the bulk aspect. 
As discussed in Ref.~\cite{Izumi2007}, it may be out of applicable range of the DGP-braneworld description because 
both contributions diverge. However, the tensionless solution sets the expectation that even in an UV completion 
for the DGP model less matter field is to construct the solution with two branes. 

In general, the existence of the configuration founded here could lead to the semiclassical instability of 
DGP braneworld. 
If so, this may be fatal to the DGP braneworld model.
But, as stressed in Ref. \cite{Witten}, the supersymmetry may protect such instability. 
Moreover, there is a question for the initial state before the decay; that is, is the DGP vacuum with the single brane  the initial state for the solution founded here? 
Since the size of the junction point is larger than $r_i$, the bubble size is also large. 
$r_i$ is expected to be a cosmological scale, and then 
the decay rate of the DGP vacuum to the bubble is exponentially 
suppressed. This is because the decay to spacetimes with large volume has a tendency to be suppressed as usual. 
The solutions constructed in this paper have the same asymptotic structure as the normal branch solutions 
on the Minkowski bulk with compactification to the extra direction. 
Thus, the solutions probably describe the spacetime after the decay of the normal branch. 
However, if one could have the solutions for the self-accelerating branch, the 
suppression for the decay rate to the single-brane solution might be relaxed. {The detailed analysis based 
on quantum gravity will be interesting.} 
There is also a problem that the self-accelerating brane is copiously nucleated \cite{Gregory2007}. 

{We emphasize that our solutions themselves could be worth investigating. 
The geometry on a single brane is like the Einstein-Rosen bridge. Since we consider the vacuum brane, any energy 
conditions are  not violated. This is an example of wormhole spacetime which satisfies the energy conditions. 
The detailed analysis will be reported in the near future \cite{tomikawa}.}

\begin{acknowledgments}
T. S. thanks Y. Sakakihara and Y. Yamashita for useful discussions in the early stage of this work. 
K. I. is supported by Taiwan National Science Council under Project No. NSC101-2811-M-002-103.
 {The authors would like to thank Yoshimune Tomikawa for pointing out typos.}
T. S. is supported by Grant-Aid for Scientific Research from Ministry of Education, Science,
Sports and Culture of Japan (No.~21244033 and No.~25610055). 
\end{acknowledgments}



\end{document}